# Feature Extraction for Machine Learning Based Crackle Detection in Lung Sounds from a Health Survey

Morten Grønnesby[1], Juan Carlos Aviles Solis[2], Einar Holsbø[3], Hasse Melbye[2], Lars Ailo Bongo[3]*

[1] Department of Medical Biology, UiT The Arctic University of Norway, N-9037 Tromsø, Norway.
[2] General Practice Research Unit in Tromsø, Department of Community Medicine, UiT The Arctic University of Norway, N-9037 Tromsø, Norway.
[3] Department of Computer Science, UiT The Arctic University of Norway, N-9037 Tromsø, Norway.
*Corresponding author: email: larsab@cs.uit.no; telephone: +47 920 155 08

## Abstract

**Background and Objective**: The stethoscope is a well-known and widely available diagnostic instrument. In recent years, many innovative solutions for recording and viewing sounds from a stethoscope have become available. However, to fully utilize such devices, there is a need for an automated approach for detecting abnormal lung sounds, which is better than the existing methods that typically have been developed and evaluated using a small and non-diverse dataset.

**Methods**: We propose a machine learning based approach for detecting crackles in lung sounds recorded using a stethoscope in a large health survey. Our method is trained and evaluated using 209 files with crackles classified by expert listeners. Our analysis pipeline is based on features extracted from small windows in audio files. We evaluated several feature extraction methods and classifiers. We evaluated the pipeline using a training set of 175 crackle windows and 208 normal windows. We did 100 cycles of cross validation where we shuffled training sets between cycles. For all the division between training and evaluation was 70%-30%.

**Results**: We found and evaluated a 5-dimensional vector with four features from the time domain and one from the spectrum domain. We evaluated several classifiers and found SVM with a Radial Basis Function Kernel to perform best for our 5-dimensional feature vector. Our approach had a precision of 86% and recall of 84% for classifying a crackle in a window, which is more accurate than found in studies of health personnel. The low-dimensional feature vector makes the SVM very fast. The model can be trained on a regular computer in 1.44 seconds, and 319 crackles can be classified in 1.08 seconds.

**Conclusions:** Our approach detects and visualizes individual crackles in recorded audio files. It is accurate, fast, and has low resource requirements. The approach is therefore well suited for deployment on smart devices and phones or as a web application. It can be used to train health personnel or as part of a smartphone application for Bluetooth stethoscopes.

# 1 Introduction

The stethoscope is a well-known and widely available diagnostic instrument. Health care personnel routinely use it to listen for abnormal sounds in the lungs to establish a diagnosis. Even though lung auscultation is a very old technique, recent technological advances in the fields of hardware, acoustics, and digital sound analysis and classification gives new possibilities that should be further explored. This is reflected by the many new commercial solutions for recording and viewing sounds from a stethoscope, such as the Bluetooth solution for smartphones from Eko Devices (https://ekodevices.com/), and the MIT mobile stethoscope [1]. However, these solutions do not have automated approaches for detecting abnormal sounds in lung sounds, that can easily be integrated with smart devices and phones.

Crackles are short, explosive nonmusical sounds heard mostly during inspiration [2]. These sounds are present in lung and heart related diseases like chronic obstructive pulmonary disease (COPD), pneumonia, heart failure, and asbestosis. In these diseases, the presence of crackles helps to stablish a diagnosis [3]–[5]. These diseases represent a major public health problem. In 2012, over 3 million deaths were caused by COPD which represented a 6% of the total deaths in that year [6]. Recent reports [7] state that 23 million people worldwide have a diagnosis of heart failure.

The use of lung auscultation in the diagnosis and treatment of disease has been questioned lately (http://www.npr.org/sections/health-shots/2016/02/26/467212821/the-stethoscope-timeless-tool-or-outdated-relic, http://www.telegraph.co.uk/news/health/news/10592653/Stethoscopes-on-their-way-out.html). This, due to concerns about the subjectivity of the technique and the introduction of new diagnostic technologies like ultrasound and CT-scans. However, the use of lung sounds in the investigation of disease has advantages in terms of costs and availability. In addition, lung sounds have the ability to reflect rapid changes, and are therefore useful in evaluating treatment responses, home monitoring, and maybe predicting exacerbations of disease [8]–[10].

Detecting crackles in lung auscultation is challenging for two reasons. First, the (crackle) signal to noise ratio is low in a sound file since crackles have a short, 5-40ms, duration. Second, other sounds are very similar, such as the stethoscope touching clothing or chest hair. Current approaches for automatic detection of crackles in lung sounds have shown promise and they have achieved high specificity and sensitivity for small test data ([11], [12] provides reviews, CORSA [13] recommends standard for terms and techniques). Most are rule based [14], [15], and hence detect crackles using a set of predefined parameters that have been extracted from a small set of sample audio files using signal processing techniques. Recently several machine learning based approaches, based on for example SVMs and Neural Networks [16]–[20] have been introduced. These have the advantage over rule based methods that the classification rules are automatically learned from the dataset. However, these have also been trained using small datasets, often with cases selected among patients with known lung diseases.

Machine learning based classification such as automatic speech recognition [21] and automatic acoustic event detection [22] is a very active research field with widely used solutions. However, such generic approaches are not well suited to detect abnormal sounds such as crackles, since these are considered as noise to be ignored by these applications.

Here we describe our machine learning based approach for automatic crackle classification in lung sounds. We train our model using lung sounds recorded in a large health survey that represents the general population. This differs from earlier work, where the lung sounds are often from a small number of patients with lung diseases. These features are fast to analyze and hence they can be used to detect abnormal lung sounds in real time, for example during an auscultation.

To train and evaluate our classifier, we selected 209 files classified by expert physicians to contain crackles from a large reference database with 36054 sounds recorded for 6009 people as part of a large health survey (Tromsøundersøkelsen 7, https://uit.no/forskning/forskningsgrupper/sub?sub_id=503778&p_document_id=367276). Our analysis pipeline is based on features extracted from small windows in these files. The approach is cheap, easy, and convenient to use. It only requires a stethoscope with a microphone for recording sounds, and the sounds can be recorded in a clinical setting with background noise. The sounds are uploaded to our server for analysis. We present the results in a web application, using a visualization that shows the detected crackles in a sound recording, and associated confidence scores for the crackles. The visualization is portable, and the results are displayed using web technology so they can be viewed on mobile phones, tablets, and PCs.

## 2 Methods

Our approach provides automatic crackle detection and annotation in a user environment that is readily accessible to health personnel and other users. A user records sounds using a stethoscope with a microphone, and then uploads these to our server for classification and annotation (Figure 1). The results are presented in a web application that provides an interactive visualization for the end users. In addition, the results can be exported as Excel or comma separated tabular text files.

We use supervised learning to classify sounds as either crackles or normal (not-crackles). Such a machine learning method can classify big datasets without intervention from a human expert. Our supervised approach requires prior knowledge in the form of a pre-classified training set with crackles and normal sounds. We select crackle specific features using sounds from a large reference database with expert classified lung sound recordings. The patterns in these are then summarized as a feature set. The features are used to train a model used to classify sounds as either normal or crackles. There are several methods for feature extraction, and several classifiers with associated learning methods. In addition, an important challenge when developing a classification approach is to find and optimize the methods that works best for a dataset. In this paper, we present and evaluate several feature extraction methods and classifiers.

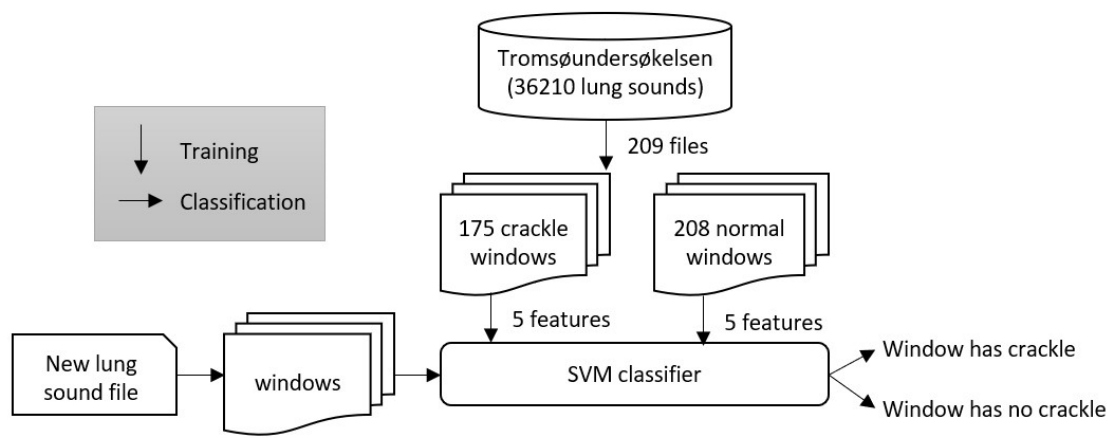

*Figure 1:Training data processing (top-down) and classification workflow (left-right).*

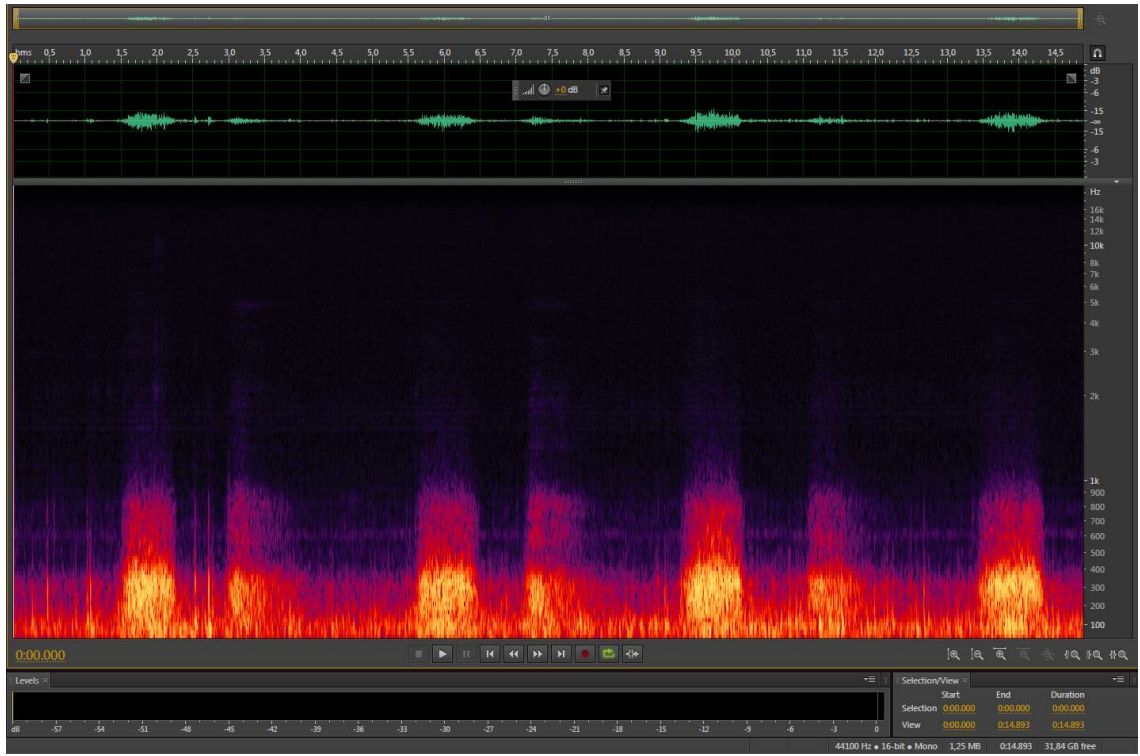

*Figure 2: Screenshot from Adobe Audition 5.0 used by the experts to classify the lung sounds in our reference database.*

## 2.1 Data acquisition

We used a sample of sound files from adults participating in the Tromsø 7 study. The Tromsø study is an epidemiological prospective study of health conditions and chronic diseases. To investigate the validity of pulmonary auscultation as a diagnostic method, we recorded lung sounds using an electret microphone (MKE 2-eW Gold, Sennheiser electronic GmbH & Co. KG) inserted at the tube of a stethoscope, 10 cm away from the chest piece. The microphone was tuned to a sensitivity of -12 dB to reduce crackle like artifacts. For the first 300 persons, we used a cardiology stethoscope (Littman Cardiology II, 3M corporation) and for the rest of the participants we used a different model (Littman Classic II SE, 3M corporation). The reason to change stethoscope was a better performance in terms of reduced low frequency noise. The sound files were captured in Wave (.wav) format at 44.100 Hz sampling rate. We have not processed the WAV files after recording. We asked the patients to breathe in and out with an open mouth and deeper than normal. We recorded in six different locations in the thorax for a period of 15 seconds in each case. In total we recorded 36210 sounds from 6035 individuals. From these 45.2% were male. Because of the high attendance rate of the study (65%) we believe that the random sample we took is representative for its age group in the area. Therefore, the prevalence of heart and lung diseases are similar to that reported in the general population [23], [24].

## 2.2 Expert classified reference database

We have started creating a reference database for lung sounds. The recordings are classified at two levels. First, two observers independently classified each recording using Adobe Audition 5.0 to listen to the lungs sounds and inspect spectrogram visualizations (Figure 2). The classification scheme had the following variables:

1. Abnormal sound

2. Inspiratory wheeze
3. Expiratory wheeze
4. Inspiratory crackle
5. Expiratory crackle
6. Other abnormal sound
7. Not classifiable

If there were any disagreements at the first step, the recordings were discussed in a meeting between the 2 observers and a third expert on lung sounds. After a discussion, the final decision agreed on, in a few was cases after voting. This dataset is multi label, so a sound file can contain both crackles and wheezes. At the time of writing we have classified 8784 files, of which 333 have crackles (3.8%).

## 2.3 Manually created training sets with crackle and normal windows

To train the classifier we used the first 209 crackle files that were classified as either inspiratory or expiratory crackles. The sounds already classified as containing crackles were plotted as waveforms using Adobe Audition 5.0, and then one of the authors visually identified crackles in the waveform and latter verified these by listening to the selected part of the file. Even though the actual windows were not verified by other experts the whole recording was evaluated and validated by at least two experts, so we believe most of these represents actual crackles.

For each crackle, we record the approximate start and end time. In total, we used 175 crackles as our training set. We also randomly selected 208 parts that do not contain crackles from the same 209 crackles files. These parts represent normal sounds.

## 2.4 Preprocessing: split file into windows

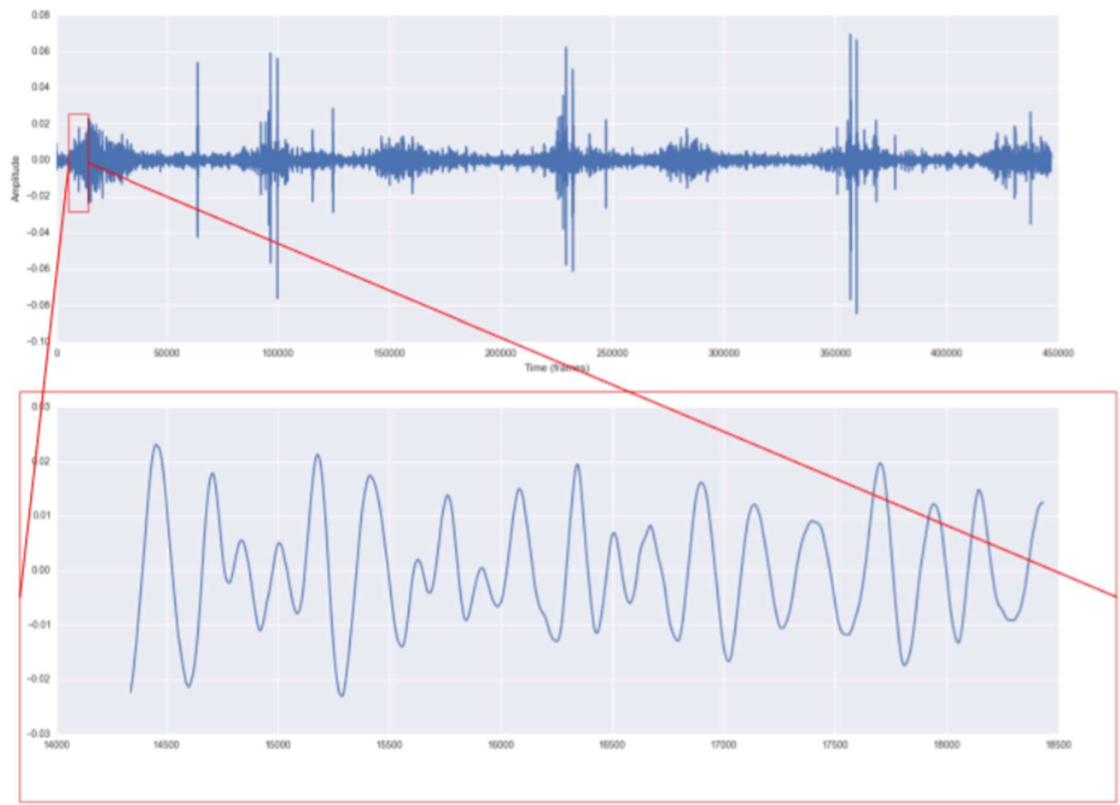

*Figure 3: The signal in audio file is divided into smaller windows. Each window is either manually examined to produce a training set, or sent to the classifier for automatic classification.*

We first split the files into 92ms windows (Figure 3). Each 92ms window contains 4096 samples, with 50% overlap between windows. The overlap ensures that a crackle is not split between two windows. The window size allows tracing abnormal sounds back to their location in time, with acceptable accuracy. Dividing the audio into windows also limits the amount of data that is analyzed at a time, which makes it easier for a machine learning algorithm to find patterns. Finally, using a fixed window as a data point ensures that a data point is not misclassified due to a lack of standardized data length and shape.

The 175 windows with crackles, and the 208 normal windows, are stored as one-dimensional arrays that contains 4096 32-bit floating point numbers. The windowing is done at analysis time, so the window size and overlap can be changed without (manually) generating a new training sets. We considered using a Butterworth bandpass filter [25] to remove frequencies above 2400 Hz and below 50 Hz since these do not contain crackle sounds. However, the Butterworth filter did not improve our results, and it did not significantly improve execution time. We therefore do not filter or do any other transformation of the sound in the windows.

## 2.5 Preprocessing: feature selection

To select the relevant features to build a model of crackles we evaluated several approaches. Our goal is to find a feature set that provides both high precision and high recall. The latter requires ignoring background noises and other additive noise such as tubing of the stethoscope that sound like crackles. We describe and evaluate in detail the best approach; a 5-dimensional vector of time domain and frequency domain features. Two, more complex, approaches are discussed in Section 4.1.

### 2.5.1 5-dimensional feature vector

We achieved the best results for a 5-dimenstional vector with four features from the time domain {*variance*, *range*, sum of simple moving average (*coarse*), sum of simple moving average (*fine*)} and one feature from the frequency domain {*spectrum mean*}. These are scaled to standardize each feature category across training observations. The advantage of using simple summary statistic features is that they are easy to relate to the actual data. The disadvantage is that a lot of information is lost using simple features.

We believe the time domain features work well with crackles due to their short-lasting explosive nature. All time domain features are calculated for the 92ms windows.

*Variance* is a measure of the spread of a distribution. It is the average of the squared deviations from the mean. Crackle windows have higher variance than normal windows due to their explosive nature, and squared errors are naturally sensitive to outliers. Normal windows may vary more in terms of zero crossing rate, but the spread is higher for crackles as they usually contain more power, or have a higher amplitude, than normal breathing.

The *range* of an audio window is the maximum value subtracted from the minimum value. Since crackles have an explosive popping noise, we believe the range of crackle windows will be wider than normal breathing. This feature is highly dependent on feature scaling, as it is highly sensitive to noise and other artefacts that may cause sudden high amplitudes in the audio.

The Simple Moving Average (SMA) gives an indication of how much the signal is changing over the course of time. We have used two different granularity levels of this feature. The *coarse* version calculates one sum for all values in an observation:

$$SMA_{coarse}(Sig) = \sum_{n=1}^{len(Sig)} |Sig_{n-1} - Sig_n|$$

The *fine* version first calculates sums for a range of windows along the signal and then selects the window with the highest amount of change:

$$SMA_{fine}(Sig) = Max(SMA_{coarse}(win_1), SMA_{coarse}(win_2) \ldots SMA_{coarse}(win_n))$$

The frequency domain feature is calculated using a one-dimensional Discrete Fourier Transform using the whole window. The complex component of the Fourier Transform is then discarded, and only the real-valued part is kept (magnitudes) and the 0-frequency component is shifted towards the center of the spectrum. After calculating the FT of the window and keeping the magnitudes, we are left with a spectrum of the distribution of different frequencies in the window.

The *spectrum mean* gives us an indication of the central tendency in the frequency domain. Crackles that occur in breathing often carry more power in higher frequencies. The center of the power distribution would naturally have a higher value for any windows containing crackles, though we have observed that this is a tendency rather than a rule.

*Feature Scaling* is necessary since audio data is non-stationary and fluctuating, so each recording might have a slightly different sound, gain, and noise. This ameliorates the effects of outliers and divergences between observations, and all features are in a standard scale compared with other observations. It is especially important with distance-based classifiers where scale is important. We do feature scaling by standardization, that is calculating and subtracting the mean, and dividing by the standard deviation of each individual feature. The standard scores (distribution mean and standard deviation) is calculated on all the features of the training set, and then applied to the test set.

## 2.6 Classifiers

We evaluated three classifiers for the 5-dimensional feature vector: SVM [26], KNN [27], and AdaBoost [28] (Decision Trees).

### 2.6.1 Support Vector Machines (SVM)

Support Vector Machines (SVMs) are popular and widely used classifiers, and an SVM also performed best on our features. The SVM separate our two classes, crackles and normal windows, using a hyperplane that maximize separation between observations of the two classes. The shape of the hyperplane is determined by a *kernel function.* We achieved the highest classification accuracy using the Radial Basis Function Kernel. We find the positive constant *C,* which controls the influence or cost of misclassification, using *Grid Search* that fit different values for *C* to different classifiers and then selects the highest scoring classifier.

### 2.6.2 K-Nearest Neighbor (KNN)

The K-nearest Neighbors (KNN) method is a non-parametric, lazy method, that does not make any assumptions about the structure of the underlying data, and it does not require a training step. Class membership of an unseen data point is determined by the *k* closest training observations in the feature space. We select *k* using grid search. We found that a small *k* (between 2-4) gives distinct boundaries between two classes that have, as in our case, small margins. We found that Euclidian distance performed best for our data.

We also evaluated *dynamic time warping* since it can work better for signals that differ in time and speed, such as crackles. However, for our summary 5-dimensional feature vectors dynamic time warping does not perform better than Euclidian distance. In addition, it significantly increases the classification time.

### 2.6.3 Adaptive Boosting and Decision Trees

Adaptive Boosting, or AdaBoost, is a meta-classifier that uses a collection of classifiers of the same type. The individual classifiers do not need to perform exceedingly well, if their prediction is better than random guess (error rate smaller than 0.5). The idea is to iteratively train classifiers that focus on the observations where previous classifiers went wrong by weighting these misclassified observations more heavily than the correctly classified observations. The output of the algorithm is a weighted sum of all the classifiers. Each classifier is weighted based on its error rate.

## 2.7 Server implementation

The server is implemented in Python 2.7 using Scikit Learn [29]. We use Python and Sklearn due to its flexibility and ease of use. The server is portable across different operating systems. Our pipeline is single-threaded. Due to the low execution time of both training and classification, we did not use optimized libraries or parallel execution.

## 2.8 Evaluation methodology

Each of the features was tested by running a train-validate cycle 100 times, and then averaging the F1-score across all cycles. The train-validate split is done on the training set, consisting of 175 crackle windows and 208 normal windows. Each cycle splits the training set into 70% training, which is used for training and grid search parameter tuning, and 30% validation. We report the precision (positive predictive value), recall (true positive rate), and F-1 score (harmonic mean between the two preceding measurements).

To measure the performance of our server we ran our server using cProfiler for Python 2.7.9 on Windows 10 Pro 64-bit, on a machine with Intel Core-i5-4570s, with four 2.90GHz cores and 6GB of DRAM.

# 3 Results

Our evaluation results provide answers to the following three questions:

1. How well do each of the five features individually separate between crackles and normal windows?
2. Which classifier works best for our feature vectors?
3. What is the speed-performance of our server during training and classification?

## 3.1 Feature selection

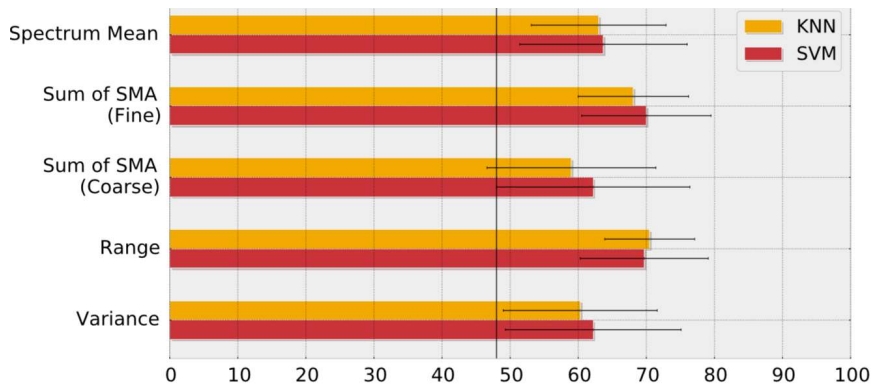

Figure 4: Average F1-score for individual features. The vertical line marks the random guess F1-Score, and the horizontal lines shows the standard deviation. The figure shows that a single feature is not good enough for crackle classification.

The univariate feature scores are between 60-70% (Figure 4). A single feature is therefore not good enough, but it is better than random guess. It should therefore be possible to combine these features to get a better separation in a higher dimensional space. The scatter matrix in Figure 5 shows the separation between normal and crackle classes. While there is separation between classes, there is also overlap. This is also reflected in the classification results (Linear

SVM in Table 1), where we get high precision, but low recall due to the overlap of features, which are not linearly separable.

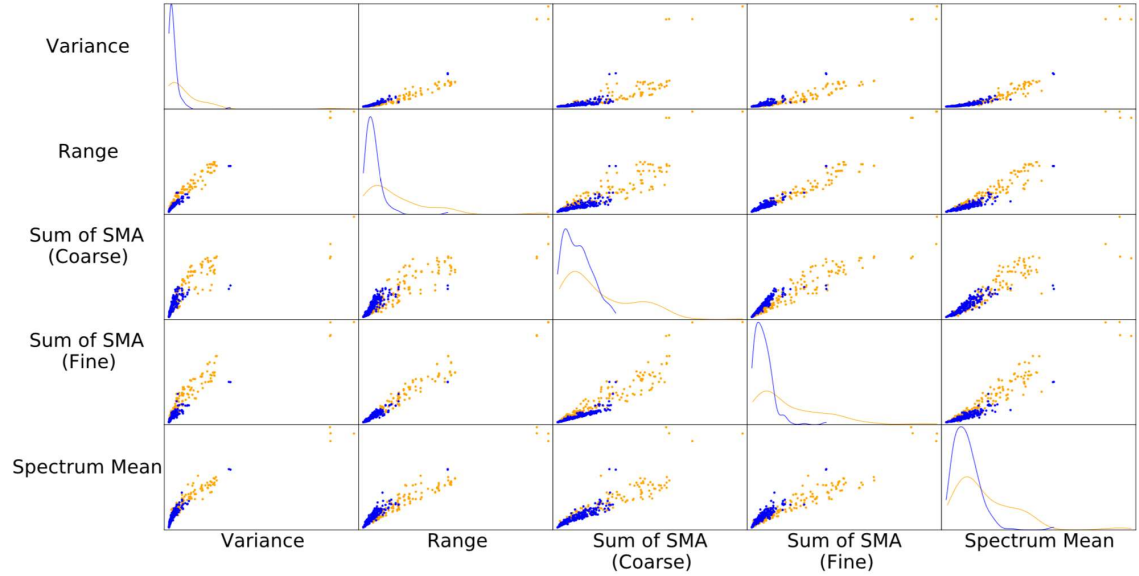

*Figure 5: Scatter matrix of the 5 feature dimensions of the training data. The yellow dots are normal observations, and the blue dots are crackles. The diagonal shows a Gaussian Kernel Density estimation. There is separation between the crackle and normal class but also overlap.*

## 3.2 Classifiers

*Table 1: Classifier performance on crackles.*

| Classifier | Precision | Recall | F1-Score |
|---|---|---|---|
| **SVM (RBF)** | 85.6 ± 6.1 | **83.6 ± 10.2** | **83.5 ± 3.6** |
| *KNN* | 84.4 ± 6.9 | 82.3 ± 11.3 | 82.5 ± 4.7 |
| *AdaBoost (Decision Tree)* | 82.7 ± 5.2 | 81.8 ± 8.3 | 81.9 ± 4.5 |
| *Linear SVM* | **88.7 ± 6.7** | 67.3 ± 8.6 | 76.0 ± 5.4 |
| *Dummy classifier (Stratified)* | 49.7 ± 7.7 | 49.7 ± 7.4 | 49.5 ± 6.8 |

SVM performs best in the cross-validation cycle results using all features (Table 1). We did a grid search for each cycle and found that a radial basis function kernel with a C parameter between 1000 and 2000 performed best. All classifiers performed better than a dummy classifier that used a stratified sampling strategy, which means that the dummy classifier chooses classes proportionally to the size of the two classes.

## 3.3 Server speed-performance

We trained the classifier in 1.44 seconds, including sequential grid search of 64 SVM parameter combinations (192 fits). Using the trained model, we classified 319 windows in 1.08 seconds. The model can therefore be used to classify crackles in real-time for example during lung auscultation.

# 4 Discussion

We found that a simple 5-dimensional feature worked best with an SVM classifier. We have also evaluated other feature extraction methods, including classifiers for these [30], [31].

The accuracy and reliability of our system is better than found in studies of health personnel [32]–[35] (although health personnel have higher accuracy than reported in [32]–[35] using our data (unpublished)). However, further research is necessary to determine the feasibility of applying our method in the clinical setting. We therefore plan to compare the performance of our system against the classifications of human experts using our full (large) data set. We also plan to compare the results of our method with other approaches for lung function evaluation such as spirometry and pulse oximetry, and to determine if our approach can improve care and health outcomes.

## 4.1 Comparison to previous results

We have not been able to directly compare our results to previous approaches since the source code and dataset are typically not available. However, our approach and results differ in several ways to the eight [36]–[43] automatic crackle detection methods in a recent review [12].

The main difference is that we use data from a health survey that represents a general population. In the previous work [36]–[43] data was from patients in hospitals. These studies had lung sound recordings from 2-36 patients with crackles, but much fewer (0-6) recordings from controls without crackles. The reported specificity and sensitivity ranges from not good ([43], 59% specificity, 24% precision) to extremely good ([38], 100% specificity, 98% sensitivity). However, these results may be due to overfitting, and the class imbalance in the training and test dataset. The results may therefore not be robust and therefore not transferable to other datasets.

Previous automatic crackle detection methods can be split (as done in [12]) into two classes: classification of single windows (segments) [36], [40]–[43] as we do, and classification of higher level events that comprise several windows [37]–[39]. It is also possible to classify at the file level.

We have two orders of magnitude more lung sounds than in [36]–[43], but we did not use more than 209 recordings in this study. From these we extracted 175 crackles and 208 normal windows and used these for training and testing. The number of windows is comparable to [36], [37], [39] that used 50-400 windows. The number of features, and type of features differ both in all previous work. We tested several of the features, some of which are descried in the negative results section below, but we did not find that these improved the accuracy of our method. Similarly, different classifiers are used. For example in [37] two classifiers that we used, kNN and SVM, are evaluated..

## 4.2 Negative results: alternative feature extraction methods and classifiers

In earlier work [30] we evaluated two alternative feature extraction methods for our data. But these did not perform as well as the simpler 5-dimensional feature described and evaluated above.

*4.2.1 Short-Time Fourier Transformation (STFT)*
We tried using the STFT spectrogram directly in an SVM classifier, but many dimensions remain after a STFT. Therefore, we only use STFT as a preprocessing step to calculate the spectrum mean of a window. One of the drawbacks of the STFT is the fixed window size, which means that there is a tradeoff between either good time resolution or good frequency resolution.

*4.2.2 Discrete Wavelet Transform*

Discrete Wavelet decomposition (DWT) is an alternative to STFT. While STFT uses fixed sized windows, the discrete wavelet transform vary the window sizes based on frequency. Higher frequencies have smaller windows, while lower frequencies have larger windows. This gives higher frequencies a better time resolution and lower frequencies a better frequency resolution. For crackles, location and duration may be important, especially in reporting and visualizing results of analyses.

*4.2.3 Spectral Flux*

Spectral flux is a measure for how quickly a signal changes over time. Spectral flux is calculated by comparing a sliding window to the previous window over a normalized waveform. We use the Euclidean Distance between the two windows. While spectral flux can be useful in onset detection, there is very little information retained from a spectral flux measurement. We believe it can be useful as a feature for detecting possible crackle candidates inside a larger audio file, but it is less useful as a feature in crackle classification.

*4.2.4 Mel Frequency Ceptrsal Coefficients*

Mel Frequency Cepstral Coefficients (MFCC) feature extraction is widely used in speech and music recognition [44]. The MFCC greatly reduces the dimensionality of the training data, but it has been designed and used for speech recognition. For crackle detection, the sound of interest is a short, explosive, non-musical sound. The MFCC represents spectral envelope of the signal, which is good for recognizing linguistic characteristics, but crackles do not adhere to these characteristics.

*4.2.5 Spectrogram Image Analysis*

An alternative approach for crackle detection is to convert the sound signal into a spectrogram and then use image analysis techniques. In [39], crackles are classified by extracting features from the elliptical pattern of a crackle in a spectrogram. We tried to replicate the results by calculating the spectrogram of a signal using the Short Time Fourier Transform (STFT) and then a histogram equalization to increase the contrast of the spectrogram. Further we used thresholding to normalize each value to either 0 or 1.

Even though we could replicate the spectrogram processing techniques, we could not accurately detect the elliptical structure of the crackle present in our spectrograms. We believe this is due our normal sounds having similar elliptical patterns that are not distinguishable from crackles. We did not have the data used in [39], so we could not do a direct comparison, so we could check if this was since we used STFT instead of the bio22 wavelet decomposition used in [39].

*4.2.6 Summary*

These alternative approaches to feature engineering provided useful classification results, but not as good as the simpler 5-dimenstional vector. We believe important challenges for crackle detection of sounds recorded in a clinical setting (or other settings such as home monitoring) includes a low signal to noise ratio, crackle-like noise artefacts, and irregular loudness. Therefore, preprocessing methods that reduce the influence of noise and better techniques for locating potential abnormal sounds in larger audio files, such as finding potential ~20ms crackles within an audio file of 15sec, are just as important as feature engineering.

## 4.3 Improvements for per file and per person crackle detection

In the above approach and evaluation, we have localized individual crackles in audio files. The methods cannot be applied directly to classify individual files and persons as either normal or having crackles for three reasons. First, our lung sound recordings are about 15 seconds and hence have about 300 windows (92ms, 50% overlap). Our methods do not have the specificity to avoid detecting at least one false positive among these windows. Second, we did not consider the natural class imbalance. Only about 5% of the recordings in our reference database (and hence population) have crackles. Training and evaluating on a synthetically balanced data set will introduce a higher false positive rate as it leads to a natural overestimation of the prior probability of a crackle. This is basically a tradeoff between false positive rate and sensitivity, as it can be hard to increase the one without also increasing the other in imbalanced data. Third, even healthy people may have some crackles. We can reduce the false positive rate by improving our methods, or by training these on a larger training set. In addition, we can use higher level information for interpretation of a file. First, the analyst may consider the number of crackles and their distribution in a recording. One, or a few, crackles in a recording is typically not of interest. Second, crackles do not occur randomly but in relation to breathing phases. Most crackles happen at an inspiratory phase and sometimes at an expiratory phase. We can therefore eliminate false positives that happen between breathing cycles. Using Parzen Windows (Kernel Density Estimation) [45], [46] to estimate the probability density function of a part of the signal, and then comparing the different estimates to find breathing phases.

### 4.4 Applications for our approach

Health workers routinely listen to lung sounds through stethoscopes during general examinations or when patients indicate respiratory distress. Such lung auscultations are an important method for physicians in decisions on treatment and referral for ultrasound or MR. However, auscultation is a subjective method and improper treatment and referrals accumulate an increased time and monetary cost. Training physicians is a challenging task because of varying perception of sound and lack of common terminology, though the latter have come more into focus for pulmonary experts. Because of these challenges, better tools for training are required and a gold standard of abnormal lung sounds is greatly needed. Training physicians using such tools, would help them to more accurately diagnose and decide a course of treatment and referral. Our approach can be used in training tools to automatically detect and highlight crackles in waveforms and spectrograms (similar to [47]). Similar visualizations are also possible to use in a smart phone application connected to a Bluetooth stethoscope as an aid for clinicians (and other people) using a stethoscope. Finally, a better set of tools for detecting abnormal lung sounds could also be used for self-monitoring of especially at home patients with chronic lung diseases.

### 4.5 Conclusion and future work

We have presented a machine learning based approach for detecting crackles in sounds recorded using a stethoscope as part of a large health survey. We evaluated several feature extraction methods, and classifiers using 209 files from a dataset with 36054 sound recordings. A simple 5-dimenstional vector and a SVM with a Radial Basis Function Kernel performed best. We achieved a precision of 86% and recall of 84% for classifying a crackle in a window, which is more accurate than found in studies of health personnel. The low-dimensional feature vector makes the SVM very fast and it can classify lung sounds in real-tome. We plan to verify that our methods work well for lung sound recordings collected using other stethoscopes, microphones, and recording environments. We believe the approach is therefore well suited for use in training of medical doctors, and for deployment on smart devices and phones. However, further research is necessary to determine the feasibility of applying our method in the clinical setting.

## 5 Acknowledgments

# 6 Conflict of interest statement

The five authors have a commercial license for the approach described and evaluated in this paper. LAB co-founder of the Medsensio AS company that provides tools for lung sound analysis.

# 7 Human rights statement

The Tromsø study was approved by the Norwegian Data Inspectorate and the Regional Ethical Committee of North Norway (REK). Only the sound files, and variables classifying the sounds were used, and identification of the participants was not possible.